# Fluidic Shaping of Freeform Optical Components

Mor Elgarisi[§], Valeri Frumkin[§], Omer Luria, and Moran Bercovici*

Faculty of Mechanical Engineering, Technion – Israel Institute of Technology

[§]Equal contribution    *Corresponding author: mberco@technion.ac.il

**Abstract**

Freeform optical components offer significant compactization of multi-lens systems, as well as advanced manipulation of light that is not possible with traditional systems. However, their fabrication relies on machining processes that are complex, time-consuming, and incompatible with rapid prototyping. This work presents the ability to shape liquid volumes and solidify them into desired freeform components, enabling rapid freeform prototyping with high surface quality. The method is based on controlling the minimum energy state of the interface between a curable optical liquid and an immersion liquid, by dictating a geometrical boundary constraint. The boundary shape is modeled as a cylinder whose arbitrary height is expressed as a Fourier series, allowing for an analytical solution of the resulting freeform surface as a sum of Fourier-Bessel functions. Each of these functions represents a different basic mode, whose superposition creates complex topographies. This solution allows deterministic design of freeform surfaces by controlling three key parameters – the volume of the optical liquid, the density of the immersion liquid, and the shape of the bounding frame. The paper describes a complete workflow for rapid prototyping of such components, and demonstrates the fabrication of a 35 mm diameter freeform component with sub-nanometer surface roughness within minutes.

## 1. Introduction

Freeform optics is a broad term that refers to any optical component, reflective or refractive, in which one or more of its optical surfaces performs complex phase manipulation on the incoming wavefront, beyond that achievable using traditional optical components. A single freeform optical component can replace the functionality of multiple traditional lenses (spherical or aspherical) within an optical system, allowing significant reduction in dimensions and assembly complexity. Moreover, freeform surfaces can provide new functionality that is not attainable using standard optics.[1–6] As reported by P.J Smilie,[7] such capabilities date back to the 1920s with the work of Kitajima who demonstrated that the respective translation of two non-spherical surfaces with respect to one another provides changes in power.[8] This concept was later expanded by Alvarez who provided a complete framework for design and construction of variable power lenses.[9]

In recent years, the advent in computer-based optical design enabled the simulation of complex components and led to tremendous growth in the use of freeform optics for new applications.[10] These include multifocal corrective eyewear,[11,12] telescopes,[13–15] beam shaping,[16] ultra-short projection lenses,[17,18] Panoramic imaging,[19] solar energy concentration[20] and photolithography.[21]

Small freeform components can be fabricated by using micro-structuring techniques such as lithography and etching.[22,23] Yet these methods are limited to characteristic scales of up to tens of microns in depth, and up to several $mm^2$ in area.[19] The fabrication of larger components relies on machining approaches such as grinding, milling, and turning, followed by polishing or finishing[10,24–27] - processes that remain complex, expensive, and time-consuming. Additive manufacturing is a natural candidate when seeking to construct arbitrary three-dimensional configurations. However, existing 3D printing technologies are primarily based on layer-by-layer fabrication and cannot yet provide the required surface quality for optical applications. Better surface quality can be obtained using post-printing processes such as reflow or coatings,[28–31] but at the cost of additional complexity, fabrication time, and reduced control over the precise shape of the surface. Furthermore, since printing time is proportional to the volume of the object (in contrast to mechanical processing that is roughly proportional to the surface area), practical considerations limit the fabrication to small lenses.[28–31] An additive manufacturing method that enables to achieve smooth optical surfaces was demonstrated by two-photon polymerization, yet this method is also limited to the scale of microns.[32,33]

Recently, work by Frumkin and Bercovici[34] suggested a new method of fabricating optical components based on shaping of liquid volumes. Their method is based on injection of a curable optical fluid into a rigid bounding frame contained within an immiscible immersion liquid environment. The authors showed that using a simple cylindrical frame, the balance between gravitational forces, hydrostatic forces, and surface tension forces, enables the creation of spherical and aspherical lenses. The method was shown to be scale-invariant, allowing the production of lenses at any size while maintaining a surface roughness of 1 nm, without the need for any further mechanical processes. Furthermore, the production time only weakly depends on the size of the produced lens.

In this work, we introduce an extension of this scale-invariant fluidic shaping method to non-axisymmetric boundary conditions, allowing for rapid fabrication of freeform optical components. We study the case of a cylindrical bounding frame whose height varies azimuthally, and based on free-energy minimization, derive an analytical model that relates the shape of this bounding frame to that of the enclosed liquid interface forming the freeform surface. These boundary conditions, which can be represented as a sum of azimuthal waves, together with the injected volume and effective buoyancy, provide an infinite number of degrees of freedom which translate into a family of freeform surfaces. These can be naturally represented by a sum of polar-Bessel functions in a similar manner to the commonly used Zernike polynomials.[35] Using bounding frames created with a standard 3D printer, we demonstrate the ease of fabrication of several freeform components. We use frames defined by a single wavenumber to demonstrate common freeform surfaces such as saddles, tilts (bifocal), and quatrefoils, as well as frames defined by superposition of waves yielding arbitrary desired surfaces. All those components enjoy a surface quality on the order of 1 nm, characteristic of the fluidic shaping method, without the need for any subsequent polishing steps.

## 2. Theory

We consider a configuration similar to that described in **Figure 1**, where an optical liquid of density $\rho$ and volume $V$ is injected into a cylindrical bounding surface of radius $R_0$ and height $h(R_0, \theta) = f(\theta)$. The liquid is suspended in an immersion liquid of density $\rho_{im}$, resulting in an effective density of $\Delta \rho = \rho - \rho_{im}$. We assume that the liquid wets the inner walls of the frame and forms an interface $h(r, \theta)$ with the immersion liquid. The shape of that interface is determined by a balance between surface tension, hydrostatic forces, and gravitational forces. The relative importance of the body force to surface forces can be expressed by the dimensionless Bond number, $Bo = \dfrac{R_0^2}{\ell_c^2} = \dfrac{|\Delta\rho| g R_0^2}{\gamma}$, where $\ell_c = \sqrt{\dfrac{\gamma}{|\Delta\rho| g}}$ is the capillary length, $\gamma$ is the interfacial energy between the two liquids, and $g$ is Earth gravity directed in the negative $\hat{z}$ direction.

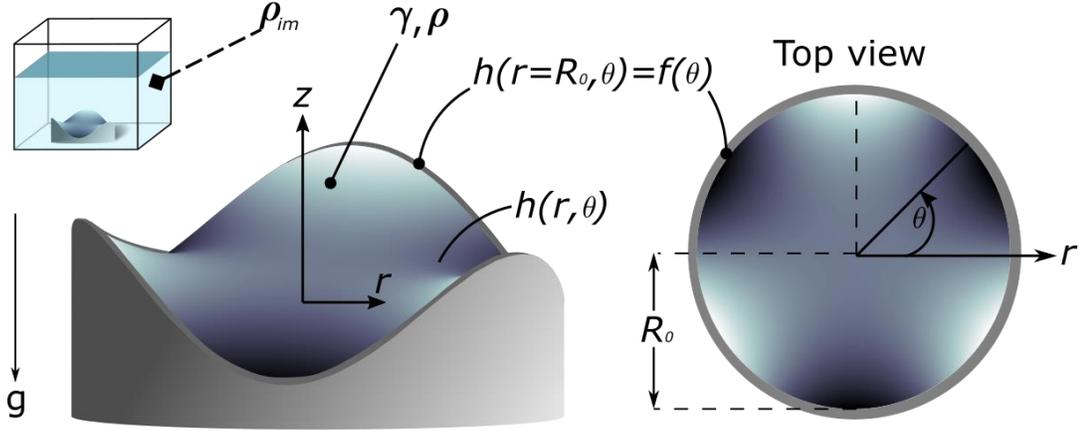

**Figure 1**. *Schematic illustration of the examined configuration, showing the coordinate system and relevant physical and geometric parameters. An optical liquid of density $\rho$ wets the inner surface of a cylindrical frame of radius $R_0$ that is entirely submerged within an immiscible liquid of density $\rho_{im}$. The surface tension between the liquids is $\gamma$. The resulting surface shape $h(r,\theta)$ is determined by a balance between gravity, hydrostatic pressure, and surface tension, subjected to the prescribed boundary condition $h(R_0,\theta) = f(\theta)$ and the total volume $V$ of the injected liquid.*

At the steady state, the fluidic interface $h(r,\theta)$ will take a shape that minimizes the free energy of the system under a fixed volume constraint. The free energy functional, $\Pi$, is given by

$$\Pi = \int_0^{2\pi}\int_0^{R_0} F(r,\theta)\,dr\,d\theta; \quad F(r,\theta) = \left(\gamma\sqrt{1+\left(\dfrac{dh}{dr}\right)^2 + \dfrac{1}{r^2}\left(\dfrac{dh}{d\theta}\right)^2} + \dfrac{1}{2}\Delta\rho g h^2 + \lambda h\right)r \qquad (1)$$

and is composed of two contributions: the surface energy associated with the interface between the liquids, and the gravitational potential energy that includes Earth's gravity and the hydrostatic buoyancy force. The last term in $F(r,\theta)$ represents the volume constraint, with $\lambda$ being a Lagrange multiplier.

At equilibrium, the first variation of the energy functional vanishes, i.e., $\delta\Pi=0$, yielding the standard Euler-Lagrange equation[36]

$$\frac{\partial F}{\partial h} - \frac{d}{dr}\frac{\partial F}{\partial h_r} - \frac{d}{d\theta}\frac{\partial F}{\partial h_\theta} = 0 \tag{2}$$

which can be written explicitly as

$$\left(\frac{\Delta\rho g}{\gamma}h+\frac{\lambda}{\gamma}\right)r^2 - \frac{r^2 h_{rr}\left(1+\frac{1}{r^2}h_\theta^2\right)+(rh_r+h_{\theta\theta})(1+h_r^2)-2h_\theta h_r h_{r\theta}+\frac{2}{r}h_r h_\theta^2}{\left(1+h_r^2+\frac{1}{r^2}h_\theta^2\right)^{3/2}} = 0 \tag{3}$$

where $h_r$ and $h_\theta$ are the partial derivatives of $h(r,\theta)$ with respect to $r$ and $\theta$.

We define the following dimensionless variables

$$R=r/R_0,\ \Theta=\theta,\ H(R,\Theta)=h(R,\Theta)/h_0,\ P=\frac{\lambda}{|\Delta\rho|gh_0},\ Bo=\frac{|\Delta\rho|gR_0^2}{\gamma},\ \varepsilon=\left(\frac{h_0}{R_0}\right)^2 \tag{4}$$

where $h_0$ is the characteristic deformation length scale, yielding the dimensionless form of equation (3)

$$(-H+P)BoR^2 - \frac{H_{RR}\left(R^2+\varepsilon H_\Theta^2\right)+(RH_R+H_{\Theta\Theta})(1+\varepsilon H_R^2)-2\varepsilon H_\Theta H_R H_{R\Theta}+\frac{2}{R}\varepsilon H_R H_\Theta^2}{\left(1+\varepsilon H_R^2+\frac{\varepsilon}{R^2}H_\Theta^2\right)^{3/2}} = 0. \tag{5}$$

For most optical elements the characteristic deformation length $h_0$ is significantly smaller than the component's radius, thus $\varepsilon=\left(\frac{h_0}{R_0}\right)^2 \ll 1$. Therefore, at the leading order in $\varepsilon$, and by defining $x=R\sqrt{Bo}$ and $H=H-P$, equation (5) reduces to

$$Hx^2 + x^2 H_{xx} + xH_x + H_{\Theta\Theta} = 0. \tag{6}$$

The general solution to this Helmholtz equation is given by[37]

$$H(x,\theta) = \sum_{n=0}^{\infty} A_n J_n(x)\cos(n\Theta) + \sum_{n=1}^{\infty} B_n J_n(x)\sin(n\Theta). \tag{7}$$

Expressing the bounding surface height as a Fourier expansion, $f(\Theta) = a_0 + \sum_{n=1}^{\infty} a_n \cos n\Theta + \sum_{n=1}^{\infty} b_n \sin n\Theta$, the constants $A_n$ and $B_n$ can be obtained by requiring equation (7) to satisfy the boundary condition $H(\sqrt{Bo},\Theta) = \dfrac{f(\Theta)}{h_0} - P$, yielding

$$H(x,\Theta) = P + \frac{(a_0/h_0) - P}{J_0(\sqrt{Bo})} J_0(x) + \sum_{n=1}^{\infty} \frac{a_n/h_0}{J_n(\sqrt{Bo})} J_n(x)\cos(n\Theta) + \sum_{n=1}^{\infty} \frac{b_n/h_0}{J_n(\sqrt{Bo})} J_n(x)\sin(n\Theta). \tag{8}$$

In dimensional form, the solution is given by

$$h(r,\theta) = \frac{a_0 - P^*}{J_0(\sqrt{Bo})} J_0\left(\frac{r}{R_0}\sqrt{Bo}\right) + P^* + \sum_{n=1}^{\infty}(a_n \cos(n\theta) + b_n \sin(n\theta)) \frac{J_n\left(\dfrac{r}{R_0}\sqrt{Bo}\right)}{J_n(\sqrt{Bo})}, \tag{9}$$

where the constant $P^*(=Ph_0)$ can be calculated via the volume constraint $V = \int_0^{R_0}\int_0^{2\pi} hr\,dr\,d\theta$, yielding

$$P^* = \frac{\sqrt{Bo}\, J_0(\sqrt{Bo})V - 2a_0 \pi R_0^2 J_1(\sqrt{Bo})}{\pi R_0^2 \left(\sqrt{Bo}\, J_0(\sqrt{Bo}) - 2J_1(\sqrt{Bo})\right)}. \tag{10}$$

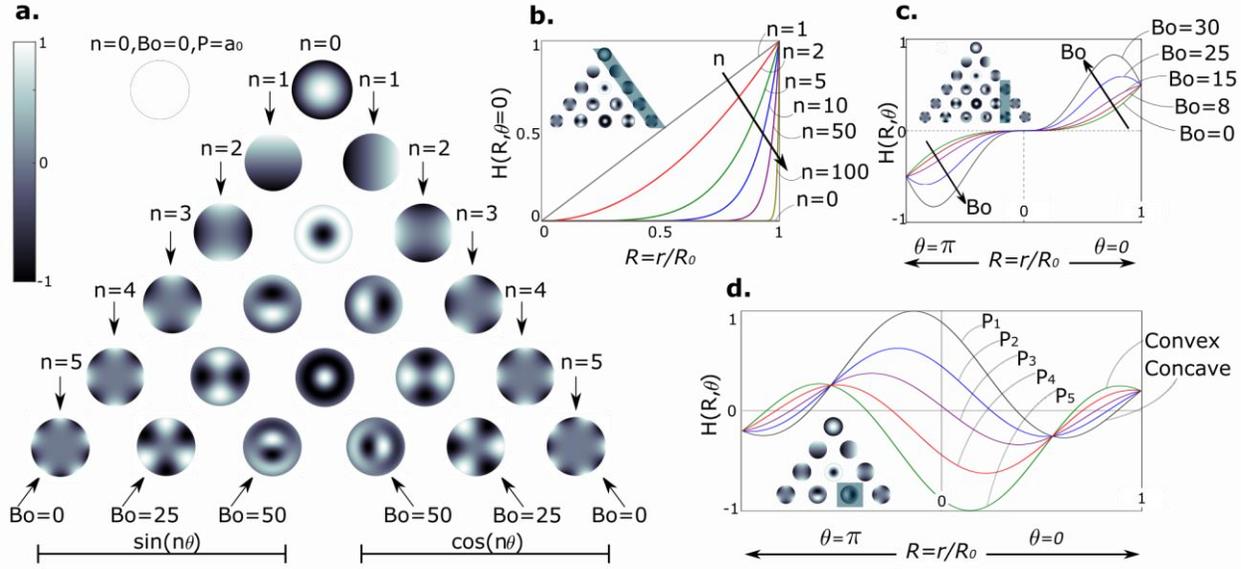

*Figure 2. Analytical results demonstrating the range of surfaces obtained by modification of the boundary conditions and the Bo number. (a) Normalized topography maps obtained for a range of Bo numbers and for periodic boundary conditions (columns) of the type $\sin(n\theta)$ (left side) and $\cos(n\theta)$ (right side). Each column corresponds to a different wavenumber $n$, and each layer in the pyramid corresponds to a different Bond number. For completeness, the trivial case of $h = 0$, obtained for $n = 0, Bo = 0$ and $V = \pi R_0^2 a_0$ (i.e. $P^* = a_0$) is presented outside of the pyramid. (b) A radial cross-section along $\theta = 0$ for $Bo = 0$, showing that an increase in the boundary's wave number, $n$, leads to faster decay of the amplitude toward the center, in accordance with equation (9). (c) A radial cross-section along $\theta = 0$ and $\theta = \pi$ for a fixed frame shape, $n = 3$, showing that increasing the Bo number produces an inflection point in the surface and shifts the maxima from the boundary inward. (d) A radial cross-section along $\theta = 0$ and $\theta = \pi$ for a fixed frame shape, $n = 1$ and a fixed Bond number, $Bo = 25$, for different P values representing different volumes, $V_{P1} < V_{P2}... < V_{P5}$. The change in volume not only affects the magnitude of the topography but can also invert the local curvature.*

**Figure 2a** presents solutions of equation (9) for several Bond numbers and several values of $n$, representing bounding frame heights with a single azimuthal frequency. We note that in contrast to $n$, which are discrete, the Bond number is continuous, and the values presented here are a subset chosen to illustrate its qualitative effect on the resulting surface. These basic modes can be conveniently arranged in a pyramid structure in accordance with the wavenumber of the periodic boundary condition, with the outermost layer of the pyramid corresponding to neutral buoyancy conditions, i.e., $Bo=0$, and an increasing Bond number towards the center. Each layer in Figure 2a, corresponding to a fixed $Bo$ number, represents a group of orthogonal surfaces that can be superposed to form a freeform surface, $h(r,\theta) = \sum_{n=0}^{\infty} h^{(n)}$, where

$$h^{(n)} = \begin{cases} \dfrac{a_n}{J_n(\sqrt{Bo})} J_n(\dfrac{r}{R_0}\sqrt{Bo})\cos(n\theta) + \dfrac{b_n}{J_n(\sqrt{Bo})} J_n(\dfrac{r}{R_0}\sqrt{Bo})\sin(n\theta) & ,n>0 \\ \dfrac{a_0 - P^*}{J_0(\sqrt{Bo})} J_0(\dfrac{r}{R_0}\sqrt{Bo}) + P^* & ,n=0 \end{cases}. \quad (11)$$

We note that $h^{(0)}$ serves as a baseline surface on which the other modes are constructed, and is also the only component in the solution that is a function of the injected volume. A particular case is that of $P^* = a_0$, (i.e. $V = \pi a_0 R_0^2$), in which the base term reduces to a constant, meaning that the base term does not contribute to optical power. For any other volume, the baseline solution represents a Bessel surface, which for the particular case of $Bo=0$ is a spherical lens. In Figure 2a the baseline surface appears at the center column of the pyramid, representing the case of a homogenous boundary condition, corresponding to a cylindrical frame with a uniform height.

For the case of $Bo=0$, the solution can be further simplified by using the limit[38],

$$\lim_{Bo \to 0} \frac{J_n(R\sqrt{Bo})}{J_n(\sqrt{Bo})} = \begin{cases} R^n, & n>0 \\ 1-R^2, & n=0 \end{cases}, \quad (12)$$

and represented conveniently using the power series

$$h(r,\theta)_{Bo=0} = a_0 - (a_0 - P^*)r^2 + \sum_{n=1}^{\infty} (a_n \cos n\theta + b_n \sin n\theta)r^n. \quad (13)$$

This representation is also equivalent to representation by Zernike polynomials of the order $Z_n^{|m|=n}$ [35]. All other surfaces in the pyramid are not equivalent to Zernike polynomials, yet as Figure 2a shows, have similar optical functionality.

Figure 2b shows that the extent to which the boundary shape affects the inner regions of the surface is a strong function of the wavenumber, $n$. As $n$ increases, the influence of the boundary is increasingly more localized, as can also be seen by the $r^n$ dependence in equation (13). This behavior holds also for other values of $Bo$. Clearly, the method cannot be used to create high-resolution features at the center of the component, but at the same time the solutions are insensitive to high-frequency deviations on the bounding frame (e.g. its surface roughness).

The Bond number represents the deviation from neutral buoyancy, and as shown in Figure 2c, tends to accentuate the effect of the bounding frame geometry. As the figure shows, increasing the Bond number can also invert the local curvature and create new local extrema. For a fixed Bond number, the injected volume can also have a significant effect on the shape of the resulting component, as illustrated by Figure 2d. By increasing the volume, the surface is transformed completely - from having a central valley with two side peaks to having a central peak with two side valleys. Combined, these three effects – the frame shape, the Bond number, and the volume, which are captured by equation (9), provide significant degrees of freedom in the design of desired surfaces.

## 3. Experimental

**Figure 3** illustrates the fabrication process of freeform surfaces using the fluidic shaping method. We use a 3D printer to print a rigid cylindrical bounding frame with a desired height variation along the azimuthal direction. We seal the bottom of the frame with a flat glass substrate and position it at the bottom of a larger container. We inject a desired volume of optical liquid into the frame and fill the container with immersion liquid until the frame is entirely submerged. At this point, additional optical liquid can be added into the frame while ensuring proper contact between the liquid and the entire inner surface of the frame. Finally, we illuminate the container with UV light for several minutes to solidify the optical liquid. The optical component is then ready and can be removed from the immersion liquid. The bounding frame can be removed from the component and reused or can remain attached to it and serve as a mechanical interface.

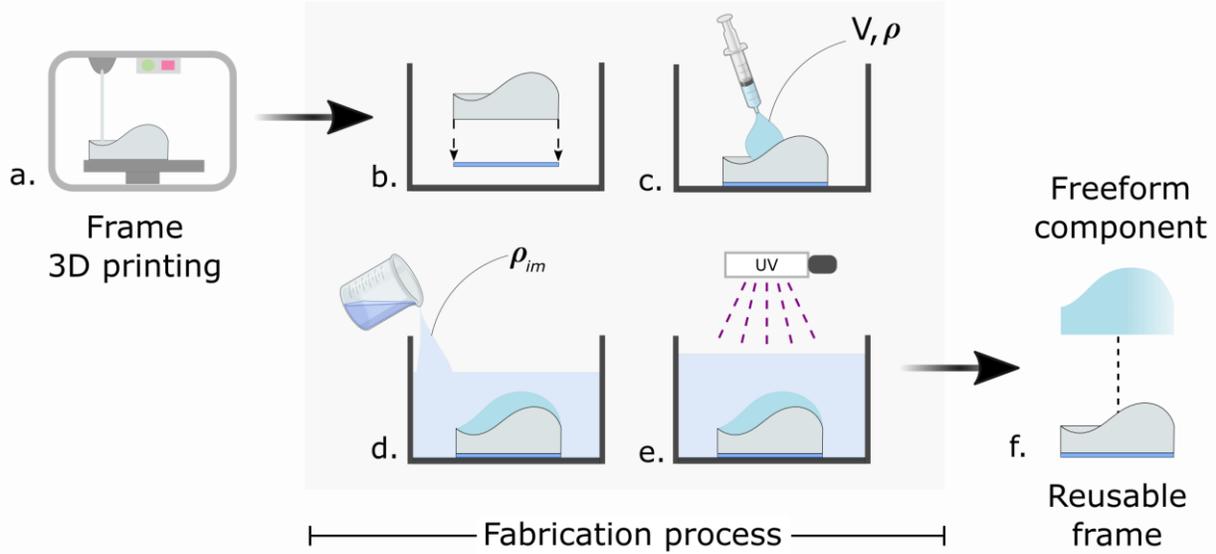

***Figure 3.*** *Illustration of the workflow for the fabrication of freeform optical components using the fluidic shaping method. (**a**) A bounding frame with a desired azimuthal height variation is printed using a 3D printer. (**b**) The frame is sealed at its bottom using a flat window and positioned at the bottom of a larger container. (**c**) The inner part of the frame is filled with an optical liquid of volume $V$, according to its design. (**d**) The container is filled with an immersion liquid of a density $\rho_{im}$ set by the desired Bond number. The immersion liquid volume is insignificant, as long as the frame and optical liquid are completely submerged. (**e**) The optical liquid is allowed to equilibrate and achieve its minimum energy state and is then illuminated with UV light to cure it. (**f**) The solid component can be removed from the immersion liquid. The frame and the immersion liquid can both be reused for the fabrication of additional components.*

**Figure 4** presents the design and fabrication of a freeform optical component using the fluidic shaping method, and a comparison of its surface topography with the theoretical prediction. Figure 4a presents the frame required for providing the boundary conditions for fabricating the component, printed on a commercial 3D printer (Form 3, FormLabs, United states). Figure 4b presents the expected resulting surface as obtained from equation (9) for a bounding frame with four sinusoidal periods, i.e. $h(R_0,\theta) = b_4 \sin(4\theta)$, where $b_4 = 0.55$ mm, $R_0 = 17.5$ mm, a Bond number of $Bo = 3$, and an injected volume of $V = 3$ ml. As described in the above procedure, we attached a flat glass window to its bottom and filled it with 3 ml of a photocurable polymer (UV resin VidaRosa J-2D-UVDJ250G, Dongguan, China) with a density of $1.07$ $g\,ml^{-1}$. To achieve the desired Bond number of 3, we filled the container with an immersion liquid composed of 26.8% glycerol in deionized water. To solidify the component, we used three 12W UV lamps with a wavelength of 365 nm, at a distance of approximately 10 cm from the component, and allowed 5 minutes for complete curing. Figure 4c presents the resulting component after solidification. The entire fabrication process required 40 minutes to complete, of which 30 minutes were spent on 3D printing of the frame, and 10 minutes on manual injection of the liquids, curing, removal from the container, and drying.

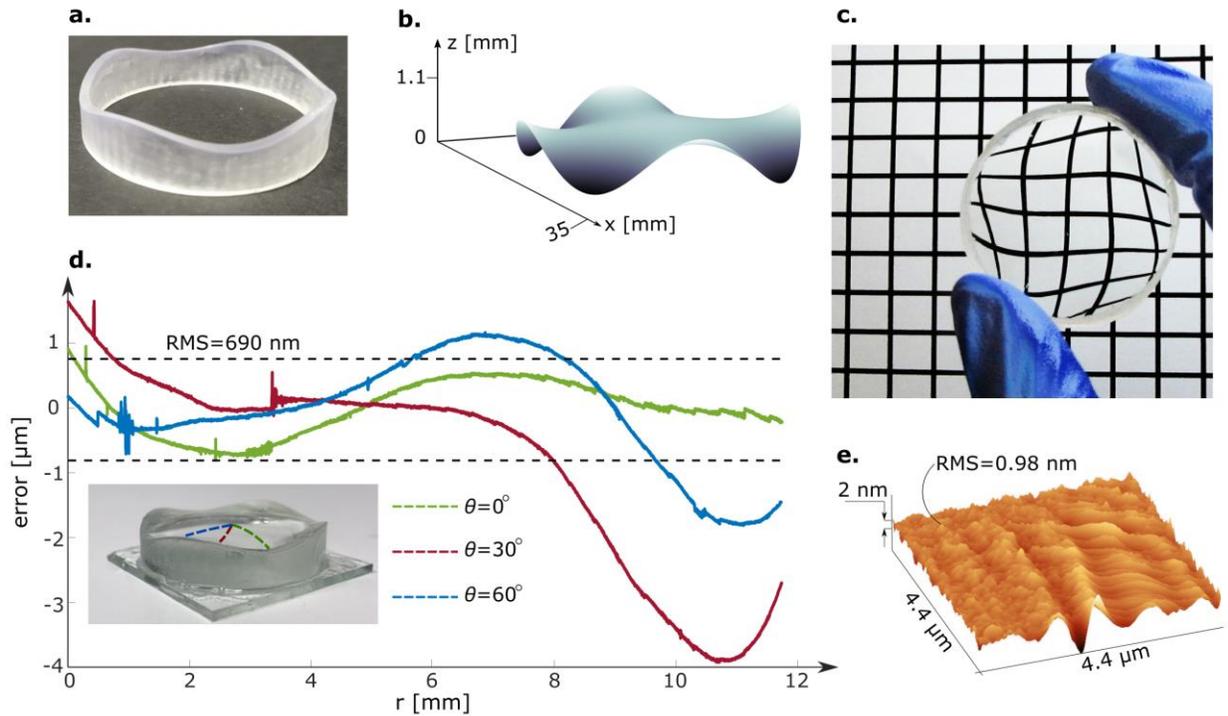

*Figure 4. Design, fabrication and characterization of a freeform optical component. (a) Image of a 35 mm diameter 3D printed bounding frame with an azimuthal height variation of $h(R_0,\theta)=0.55\sin(4\theta)$ mm. (b) Image of the predicted 3D surface, based on equation (9), for an injected volume of $V=3\,ml$ and a Bond number of $Bo=3$. (c) Image of the resulting solidified component over graph paper. (d) Plot of the error between the measured optical surface and the theoretical surface for three cross-sections $0°, 30°, 60°$. The dashed line indicates the root-mean-square of the error for the three lines, $RMS=690\,nm$. (e) An AFM measurement of surface roughness, showing an RMS of $0.98\,nm$.*

Using a digital holographic microscope (DHM R-1000 LynceeTec, Switzerland) we measured the topography of the resulting freeform surface along three radial lines ($0°, 30°, 60°$), with the angle set by a rotating stage (PR01/M, Thorlabs, New Jersey). Along each line, using a motorized X-Y stage (MS 2000, ASI, Oregon) we took a series of 40 500x500 µm images with partial overlap between them, and stitched them together to form a continuous line by cross correlation between adjacent images. Since our process is currently entirely manual, we expect inaccuracies in the injected volume and density matching compared to the desired nominal values. In addition, the 3D printed frame has a resolution of $\pm 25\,\mu m$. Hence, we first turn to extract these values as global parameters from the resulting component. To that end, we used Matlab's "fminsearch" algorithm to find the least means squares fit between the three lines and the theoretical model, with the Bond number, the injected volume, and the amplitude of the frame-height variation serving as the free parameters. The resulting values from the fit are $Bo = 2.91, V = 3.21\,ml, b_4 = 0.56\,mm$, and are within the expected error of our manual process. Figure 4d presents the difference between the measured surface height along the three radial lines, and the theoretical values along these lines using the extracted parameters. The results show that for radii smaller than 8 mm the differences between theory and measurements are capped at approximately 1 µm, and grow to 4 µm toward the edges of the frame. The RMS of the error, over all three lines, is 690 nm. Figure 4e presents the measurement of the surface roughness of the resulting component over an area of 4.4x4.4 µm, using atomic force microscopy (AFM), showing an RMS value of 0.98 nm, as expected due to the smoothness of the liquid-liquid interface.

## 4. Conclusions

We presented theory and experiments for a new method for the design and fabrication of freeform components based on shaping of fluidic interfaces. We showed that the minimum free energy state of the interface between an optical liquid and an immersion liquid is dictated by the shape of a bounding frame, the volume of the optical liquid, and the Bond number representing the difference in their densities. We presented a general solution for the case of a cylindrical bounding frame, in which the surface can be represented by a sum of Fourier-Bessel functions, which are known to have similar functionalities for optical application as other commonly-used functions.[39,40] However, summation is allowed only over functions sharing a fixed Bond number, thus limiting the range of surfaces that can be fabricated in practice. On the other hand, traditional machining approaches present their own set of constraints on the surfaces that could be produced,[1,2] whereas the liquid shaping approach guarantees that any mathematically allowed surface can in fact be manufactured with an appropriate frame. The fabrication time, dominated by the 3D printing of the frame, is on the order of tens of minutes and is independent of the complexity of the frame shape. Owing to the natural smoothness of liquid-liquid interfaces, the resulting surface roughness is on the order of 1 nm without the need for any post-polishing steps.

In this work, we limited our analysis to elements with a single freeform surface, i.e. plano-freeform components. However, the method can be directly expanded to create components with freeform surfaces on both sides. This could be achieved using a frame that is varying azimuthally on both ends and suspended in the container such that the optical liquid injected into the frame has two contact surfaces with the immersion liquid. Moreover, enclosing each end of the frame within a different immersion liquid would result in surfaces that are based on different Bond numbers, allowing to increase the degrees of freedom of the component.

We further limited our analysis to the case of small values of $\varepsilon = (h_0 / R_0)^2$, allowing to linearize the minimum energy equation and obtain an analytical solution. While this assumption holds very well for the majority of optical surfaces of interest, solving for surfaces with larger amplitude variations would still be possible by solving numerically the full nonlinear equation (3). Another interesting expansion, which may add additional degrees of freedom to the surfaces that can be produced, is the use of arbitrary (i.e. non-cylindrical) bounding frame shapes. Finally, while our work here focused on fabrication of solidified components, the optical fluid may purposely remain in its fluidic state, allowing the implementation of dynamically controlled optical components. Simple real-time changes may be possible by injecting or aspirating optical liquid, or by adjusting the immersion liquid density. A higher level of control could perhaps be achieved by implementing a deformable bounding frame whose shape can be modified in time.

**Acknowledgements**

We thank Baruch Rofman for performing the AFM surface quality measurements, and Rishabh Das for critically reviewing an earlier version of the derivation. This project has received funding from the European Research Council under the European Union's Horizon 2020 Research and Innovation Programme, grant agreement 678734 (MetamorphChip).

10. J. Ye, L. Chen, X. Li, Q. Yuan, Z. Gao, *Opt Eng*. **2017**, 56(11), 1.

11. A. Cao, J. Wang, H. Pang, M. Zhang, L. Shi, Q. Deng, S. Hu, *Bioinspir Biomim*. **2018**, 13(2), 026012.

12. W. Jia, B. Zhang, S. Li, *Int J Opt*. **2019**, 2019, 1-12.

13. B.V. Rao, K.V. Sriram, C.S. Narayanamurthy, *J Astron Telesc Instrum Syst*. **2021**, 7(01).

14. N. Bregenzer, M. Bawart, S. Bernet, *Opt Express*. **2020**, 28(3), 3258.

15. S. Bernet, *Appl Opt*. **2018**, 57(27), 8087.

16. Z. Feng, L. Haung, M. Gong, G. Jin, *Optics Express*. **2013**, 21, 12.

17. B. Yang, K. Lu, W. Zhang, F. Dai, presented at SPIE Opt Eng and App, San Diego, California, USA, September, **2011**.

18. Z. Zhuang, Y. Chen, F. Yu, X. Sun, *Applied optics*. **2014**, 53, 22.

19. T. Ma, J. Yu, P. Liang, C. Wang, *Opt Express*. **2011**, 19(5), 3843.

20. P. Zamora, A. Cvetkovic, M. Buljan, M. Hernández, P. Benítez, J. C. Miñano, O. Dross, R. Alvarez, A. Santamaría, IEEE 34th PVSC, Philadelphia, PA, USA, June, **2009**.

21. R. Wu, H. Li, Z. Zheng, X. Liu, *Appl Opt*. **2011**, 50(5), 725.

22. F. Fang, N. Zhang, X. Zhang, *Adv Opt Technol*. **2016**, 5(4).

23. M. Roeder, T. Guenther, A. Zimmermann, *Micromachines*. **2019**, 10(4), 233.

24. L. Zhu, Z. Li, F. Fang, S. Huang, X. Zhang, *Int J Adv Manuf Technol*. **2018**, 95(5-8), 2071-2092.

25. Z. Xia, F. Fang, E. Ahearne, M. Tao, *J Mater Process Technol*. **2020**, 286, 116828.

26. V. Mishra, D. Sabui, D.R. Burada, V. Karar, S. Jha, G.S. Khan, *Mater Manuf Process*. **2020**, 35(7), 797-810.